\documentclass[aps,prl,reprint,onecolumn,12pt,superscriptaddress]{revtex4-1}
\usepackage{graphicx}%
\usepackage{float}
\usepackage[T1]{fontenc}
\begin{document}
\title{Giant Zeeman splitting inducing near-unity valley polarization in van der Waals heterostructures}


\author{Philipp Nagler\footnote{philipp.nagler@ur.de}}
\affiliation{Institut f\"ur Experimentelle und Angewandte Physik,
	Universit\"at Regensburg, D-93040 Regensburg, Germany}
\author{Mariana V. Ballottin}
\affiliation{High Field Magnet Laboratory (HFML - EMFL), Radboud University, 6525 ED Nijmegen, The Netherlands}
\author{Anatolie A. Mitioglu}
\affiliation{High Field Magnet Laboratory (HFML - EMFL), Radboud University, 6525 ED Nijmegen, The Netherlands}
\author{Fabian Mooshammer}
\affiliation{Institut f\"ur Experimentelle und Angewandte Physik,
	Universit\"at Regensburg, D-93040 Regensburg, Germany}
\author{Nicola Paradiso}
\affiliation{Institut f\"ur Experimentelle und Angewandte Physik,
	Universit\"at Regensburg, D-93040 Regensburg, Germany}
\author{Christoph Strunk}
\affiliation{Institut f\"ur Experimentelle und Angewandte Physik,
	Universit\"at Regensburg, D-93040 Regensburg, Germany}
\author{Rupert Huber}
\affiliation{Institut f\"ur Experimentelle und Angewandte Physik,
	Universit\"at Regensburg, D-93040 Regensburg, Germany}
\author{Alexey Chernikov}
\affiliation{Institut f\"ur Experimentelle und Angewandte Physik,
	Universit\"at Regensburg, D-93040 Regensburg, Germany}
\author{Peter C. M. Christianen}
\affiliation{High Field Magnet Laboratory (HFML - EMFL), Radboud University, 6525 ED Nijmegen, The Netherlands}
\author{Christian Sch\"uller}
\affiliation{Institut f\"ur Experimentelle und Angewandte Physik,
	Universit\"at Regensburg, D-93040 Regensburg, Germany}
\author{Tobias Korn\footnote{tobias.korn@ur.de}}
\affiliation{Institut f\"ur Experimentelle und Angewandte Physik,
	Universit\"at Regensburg, D-93040 Regensburg, Germany}
\email{tobias.korn@physik.uni-regensburg.de}

\begin{abstract}
\textbf{Monolayers of semiconducting transition metal dichalcogenides exhibit intriguing fundamental physics of strongly coupled spin and valley degrees of freedom for charge carriers \cite{Xiao2012,Mak2012a,Zeng2012a,Xu2014}. While the possibility of exploiting these properties for information processing stimulated concerted research activities towards the concept of valleytronics \cite{Schaibley2016}, maintaining control over spin-valley polarization proved challenging in individual monolayers. A promising alternative route explores type II band alignment in artificial van der Waals heterostructures \cite{Geim2014}. The resulting formation of interlayer excitons combines the advantages of long carrier lifetimes and spin-valley locking \cite{Kang2013,Kosmider2013,Fang2014,Lee2014,Rivera2015,Rivera2016}. Here, we demonstrate direct magnetic manipulation of valley polarization in a WSe$_2$/MoSe$_2$ heterostructure through giant valley Zeeman splitting of interlayer transitions.
Remarkably, even after non-selective injection, the observed $g$ factor as large as $-15$ induces near-unity polarization of long-lived excitons with 100\,ns lifetimes under magnetic fields. The demonstrated control of the spin-valley physics highlights the exceptional aspects of novel, artificially designed material systems and their promise for atomically-thin valleytronic devices.}
\end{abstract}

\maketitle

The materials combined in the studied heterostructure are monolayers of transition metal dichalcogenides (TMDCs) where MX$_2$ denotes M=Mo, W and X=S, Se, Te. These systems were shown to host direct optical transitions in the visible spectral range at two inequivalent valleys in momentum space, labeled K$+$ and K$-$, which are situated at the corners of the hexagonal Brillouin zone \cite{Xu2014}.
Since spin and valley of charge carriers are coupled at the K points due to the broken inversion symmetry in the monolayer combined with strong spin-orbit coupling, it is possible to selectively address and read out the valley index optically by helicity-resolved measurements \cite{Mak2012a,Zeng2012a}. However, as the two valleys are linked by time-reversal symmetry, external fields are required to break their energy degeneracy, an issue of central importance for future valleytronic devices. Recently, the effective manipulation of the valley pseudospin energy has been demonstrated for monolayer TMDCs by magnetic \cite{Aivazian2014,Li2014,Srivastava2015,Macneill2015a,Wang2015,Stier2016,Plechinger2016} and electric fields \cite{Sie2014,Kim2014}. However, the extremely short lifetimes of excitons \cite{Poellmann2015,Robert2016} and the fast polarization dephasing mechanisms \cite{Glazov2014} render the implementation of individual TMDC monolayers challenging for valleytronics.

At the same time, the rapid development of transfer techniques has opened up a vast parameter space of artificial van der Waals heterostructures, where different two-dimensional (2D) materials are deterministically stacked upon each other. For TMDCs, the resulting type II band alignment and subsequent rapid charge transfer \cite{Hong2014} leads to the formation of interlayer excitons (IEX), where electrons and holes are situated in different layers.
For stacking angles close to 0 and 60 degrees, negligible momentum mismatch allows radiative recombination of charge carriers at the K$+$ and K$-$ points \cite{Yu2015a}, leading to pronounced light emission below the energies of the individual monolayer transitions. Nevertheless, due to the spatial separation, the wavefunction overlap in the out-of-plane direction is reduced, facilitating long carrier lifetimes of the interlayer excitons in 2D heterostructures \cite{Rivera2015}, further supplemented by the possibility of spin-valley injection in close analogy to the monolayer systems \cite{Rivera2016}. Most importantly, as we demonstrate in the following, the specific alignment of atomically-thin layers in a heterostructure leads to intriguing fundamental physics, exclusive for such artificial systems and allowing for highly efficient external manipulation of the spin-valley degrees of freedom. In particular, we show that the magnetic coupling of electronic transitions can be strongly enhanced in TMDC heterobilayers, exhibiting a giant valley Zeeman splitting with an effective $g$ factor of about $-15$, that exceeds typical values for both TMDC monolayers and more conventional nonmagnetic semiconductor heterostructures by far. Of central consequence is the resulting field-induced valley polarization of the long-lived charge carriers, even though both valleys in the two constituent materials are initially equally populated. The degree of polarization, reaching near-unity values arises entirely due to the strong degeneracy lifting under magnetic fields. It emerges as a major aspect of the TMDC heterostructures, advancing both our understanding of the fundamental phenomena and the application potential of pre-designed atomically-thin systems.

\begin{figure}
	\centering
	\includegraphics*[width=0.8\linewidth]{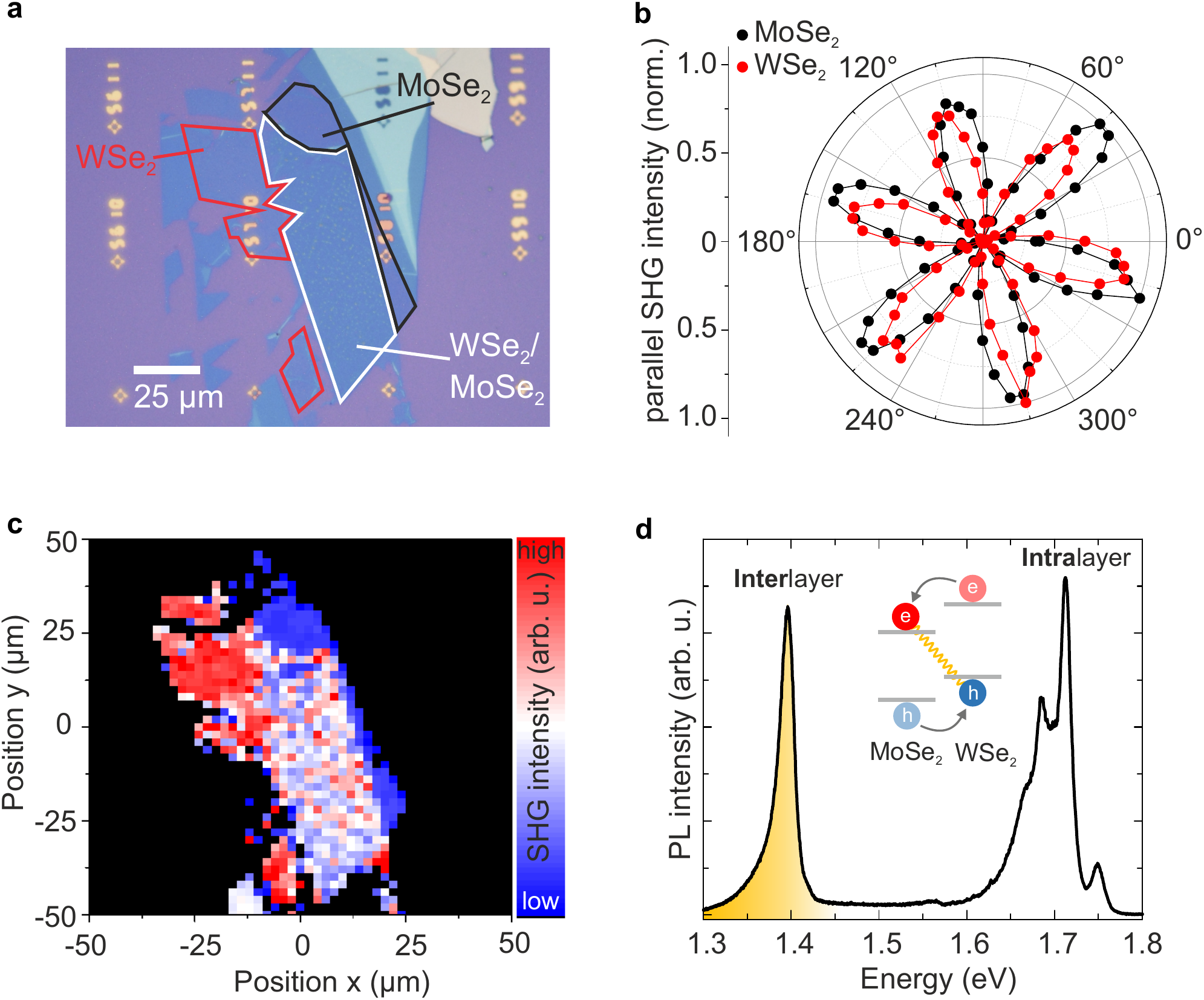}
	\caption{\textbf{Interlayer excitons in a WSe$_2$/MoSe$_2$ heterostructure with nearly $60\,^{\circ}$  angle alignment.} \textbf{a,} Optical micrograph of the WSe$_2$/MoSe$_2$ heterostructure under study. The white framed area depicts the region where the two materials overlap vertically. \textbf{b,} Angle-dependent plot of the parallel component of the SHG intensity of the individual monolayers, indicating the armchair directions of the monolayers. The relative angle between the monolayers amounts to about $54\,^{\circ}$. \textbf{c,} Spatial scan of the sample where the total SHG intensity is recorded for each datapoint. The region of the heterostructure shows clear destructive interference of the SHG signal with respect to the individual layers. \textbf{d,} PL spectrum taken on the heterostructure at 4\,K. The emission stemming from interlayer excitons is spectrally well separated from the intralayer luminescence. The inset schematically depicts the type II band alignment of the heterostructure which leads to a spatial separation of electrons and holes.
	}
	\label{Plot1}
\end{figure}

The heterostructure under study (shown in Fig.\,\ref{Plot1}a) consists of a monolayer of WSe$_2$ transferred on top of a MoSe$_2$ monolayer, exfoliated onto a SiO$_2$/Si substrate. During the transfer process, the well-cleaved axes of the monolayers are deterministically aligned parallel to each other, ensuring a twist angle of either nearly $0\,^{\circ}$ (AA-stacking) or $60\,^{\circ}$ (AB stacking) degrees. The relative angle of the stacking configuration is confirmed through spatially-resolved second harmonic generation (SHG) spectroscopy \cite{Hsu2014}. Figure\,\ref{Plot1}b illustrates the intensity of the parallel component of the SHG signal of the two individual monolayer materials in a polar plot. By fitting the data with a $\textrm{cos}^2(3\theta)$ function we directly obtain a relative stacking angle $\theta$ of either $6\,\pm1^{\circ}$  or $54\,\pm1^{\circ}$ (the two possibilities stem from the phase-insensitivity of the SHG intensity measurement on a single layer). To clarify the precise alignment, we perform a spatial scan of the resulting heterostructure where the total SHG intensity is recorded at each position, as shown in Fig.\,\ref{Plot1}c. In the overlapping region of the two layers denoted by the white framed area in Fig.\,\ref{Plot1}a, we clearly observe a pronounced destructive interference of the SHG signal with respect to the individual monolayers, consistent with a nearly $60\,^{\circ}$ stacking configuration \cite{Hsu2014}. Thus, we conclude that the sample has an AB-like stacking configuration with a relative angle of $54\,\pm1^{\circ}$.

A characteristic photoluminescence (PL) spectrum of the heterostructure recorded at 4\,K is shown in Fig.\,\ref{Plot1}d. In line with recent reports, it consists of two separate spectral regions \cite{Rivera2015,Rivera2016}: The intralayer transitions between 1.6\,eV and 1.8\,eV result from direct excitonic recombination of the constituent monolayers (WSe$_2$ and MoSe$_2$). At the same time, as depicted in the inset of Fig 1d, rapid charge transfer leads to the formation of interlayer excitons at around 1.4\,eV, which are lower in energy compared to the intralayer transitions.

\begin{figure}
\includegraphics*[width= 1\linewidth]{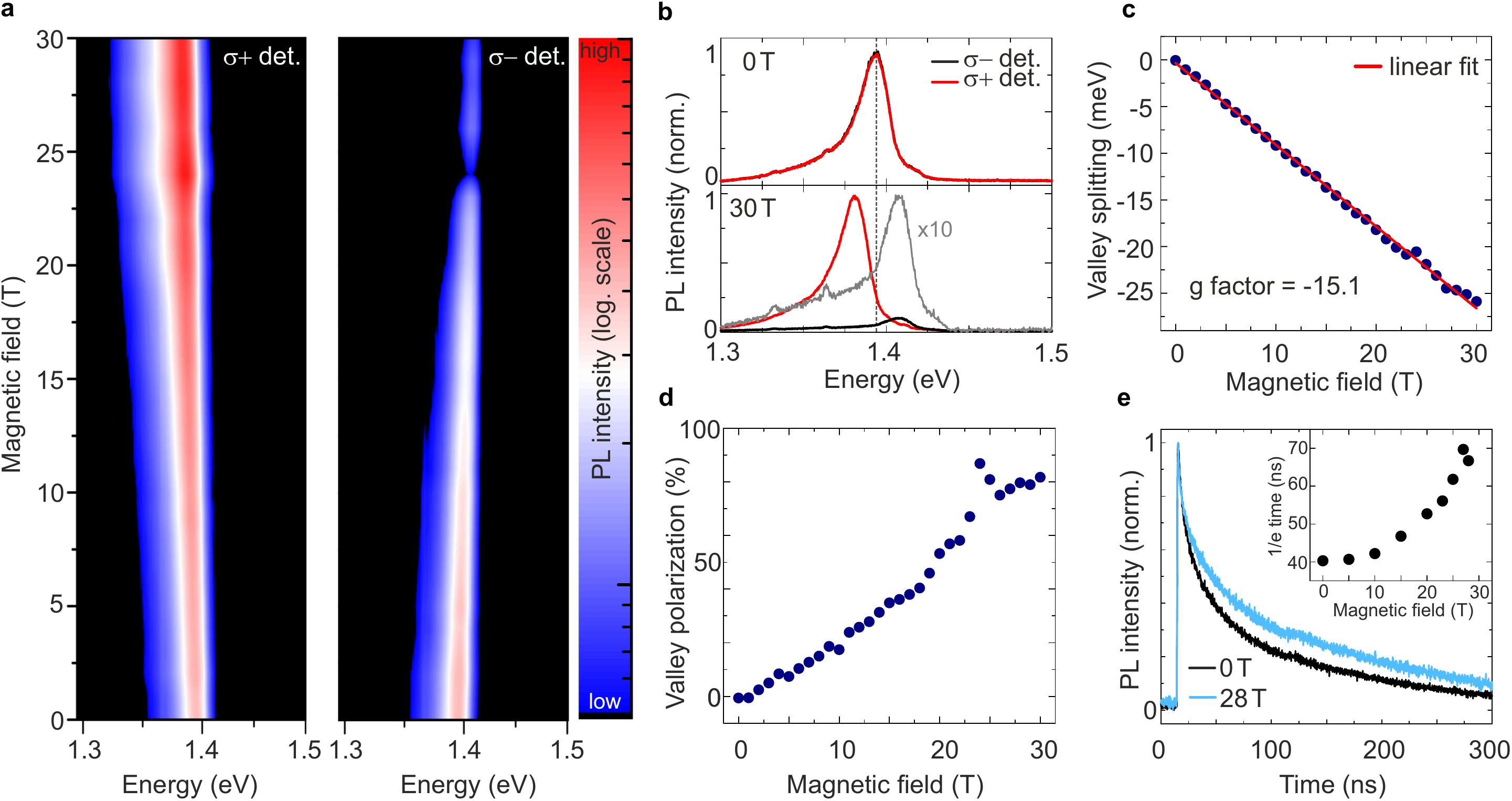}
	\caption{\textbf{Magnetic field dependence of interlayer excitons.} \textbf{a,} False color representation of the interlayer exciton PL for $\sigma+$ and $\sigma-$ polarized detection as a function of out-of-plane magnetic field up to 30\,T. The excitation is performed with linearly polarized light. For better clarity the PL intensity is plotted in logarithmic scale. \textbf{b,} Comparison of PL spectra of the interlayer exciton for 0\,T and 30\,T. At 0\,T both polarizations show the same energy and intensity. At 30\,T, the energy degeneracy is fully lifted and the emission stems almost exclusively from the $\sigma+$ transition. \textbf{c,} Corresponding valley Zeeman splitting of the interlayer exciton. The solid line corresponds to a linear fit of the data, yielding a g factor of $-15.1\pm0.1$. \textbf{d,}  Magnetic-field-induced valley polarization of the interlayer exciton. \textbf{e,} Time-resolved PL of the interlayer exciton for B\,=\,0\,T and B\,=\,28\,T. The inset shows the extracted 1/e decay-constant as function of the applied magnetic field.
	}
\label{Plot2}
\end{figure}

We now turn to the valley-resolved measurements of the interlayer transitions in an out-of-plane magnetic field (Faraday geometry). Importantly, the optical experiments are carried out by exciting the sample with linearly polarized light (laser energy 1.94\,eV), initially populating both valleys in the two monolayer constituents of the heterostructure equally. The emission is subsequently analyzed in a circularly polarized basis, allowing us to directly access the resulting valley Zeeman splitting of the transitions and quantify the degree of polarization. Figure \ref{Plot2}a shows the spectral evolution of the interlayer excitons for both detection polarizations in magnetic fields up to 30\,T. Two main observations are immediately apparent from the data: First, the peak energy degeneracy of the interlayer transitions is lifted for fields B \textgreater\,0. For rising magnetic fields, the energy of the $\sigma$+ -polarized component decreases monotonically while it increases for the $\sigma-$ -polarized component.  Second, the intensity of the interlayer exciton strongly depends on the detected polarization in the magnetic field. The $\sigma$+ and $\sigma-$ -polarized components exhibit a drastic increase and decrease in intensity, respectively, as the magnetic field is increased. These observations are further illustrated in Fig.\,\ref{Plot2}b, where the two configurations for B\,=\,0\,T and B\,=\,30\,T are directly compared. While at B\,=\,0\,T the two polarizations are of same energy and intensity, the energy splitting between the two valley configurations amounts to about 26\,meV for B\,=\,30\,T, even slightly exceeding the linewidth of the two transitions. Also, the luminescence stems almost exclusively from the $\sigma +$ transition, with the $\sigma -$ emission being strongly suppressed.

For the quantitative analysis of the data, we use Gaussian fit functions, and extract the PL peak positions of the interlayer transition for both polarizations as function of the magnetic field. The resulting valley splitting, presented in Fig.\,\ref{Plot2}c, clearly follows a linear dependence. Using the definition of the splitting as $\Delta E^{IEX}$=$E^{\sigma+}-E^{\sigma-}=g\mu_{B}B$, where $\mu_B$ is the Bohr magneton ($\approx 58 \mu eV/T$), we extract a $g$ factor of $-15.1\pm0.1$ for the interlayer exciton.
The corresponding degree of the valley polarization, defined as $P=(I_{\sigma^+}-I_{\sigma^-})/(I_{\sigma^+}+I_{\sigma^-})$, is presented in Fig.\,\ref{Plot2}d. While it is strictly zero at B\,=\,0, as expected, the emission is strongly polarized under the external magnetic field, exceeding values of 80\% for the highest fields up to 30\,T. We further note, that such overall high degree of field-induced polarization is particularly remarkable given the fact that both valleys are initially equally populated in the experiment. Using time-resolved PL measurements (see Methods) we also track the decay dynamics of the interlayer exciton in the magnetic field. The sample is excited linearly and the total PL intensity is detected. The data are presented in Fig.\,\ref{Plot2}e. The decay dynamics exhibit a complex non-exponential decay, with a 1/e time-constant of about 40\,ns at 0\,T followed by a longer decay (>100\,ns) at later times and exceeding typical values for individual TMDC monolayers by several orders of magnitude. The lifetime further increases with rising magnetic field (see inset), with the 1/e constant reaching 70\,ns at 28\,T and the longer component increasing beyond 200\,ns. Applying an external magnetic field therefore allows us not only to generate strongly valley-polarized carriers, but also to maintain the interlayer emission on very long timescales.

The determined value of about $-15$ for the $g$ factor of the interlayer exciton is in stark contrast to experimentally determined $g$ factors of excitons in individual TMDCs, found to be around $-4$ in most cases \cite{Li2014,Srivastava2015,Macneill2015a,Wang2015,Stier2016,Plechinger2016}. In monolayers, the magnitude of the Zeeman coupling is often understood in terms of a simplified semi-quantitative model, including three contributions, namely the spin, the atomic orbitals, and the valley magnetic moment \cite{Aivazian2014,Li2014,Srivastava2015,Macneill2015a,Wang2015}. Since the optical transitions are spin-conserving between conduction and valence band, the net contribution from the spin to the energy splitting of the respective resonances is zero. On the other hand, only the valence bands carry a non-zero magnetic moment $\mu_l$ from the atomic orbitals with $\mu_l=2$ for the K$+$ valley and $\mu_l=-2$ for the K$-$ valley leading to an overall splitting between the valley-selective transitions of $-4\mu_B B$. The third contribution, the valley magnetic moment $\mu_k$, arises from the self-rotation of the Bloch wavepackets \cite{Xiao2007a}. It is defined by $\pm\mu^{c}_k=m_0/m_e$ for the conduction band and $\pm\mu^{v}_v=m_0/m_h$ for the valence band in the K$+$/K$-$ valley, respectively. Assuming that optical transitions in a monolayer take place between valleys of the same index, these contributions cancel out almost entirely and the total field-induced Zeeman shift for a monolayer TMDC can then be written as $\Delta E^{1L}=E^{\sigma +}-E^{\sigma -}=-(4-2(m_0/m_e-m_0/m_h)) \mu_B B \approx -4 \mu_B B$, as $m_h \approx m_e$ in many cases. The deviations from this value are attributed both to nonequivalent effective masses of electrons and holes as well as to the complexities of the orbital contributions to the Zeeman shift beyond the simplified model \cite{Wang2015}.

\begin{figure}
	\includegraphics*[width= 0.67\linewidth]{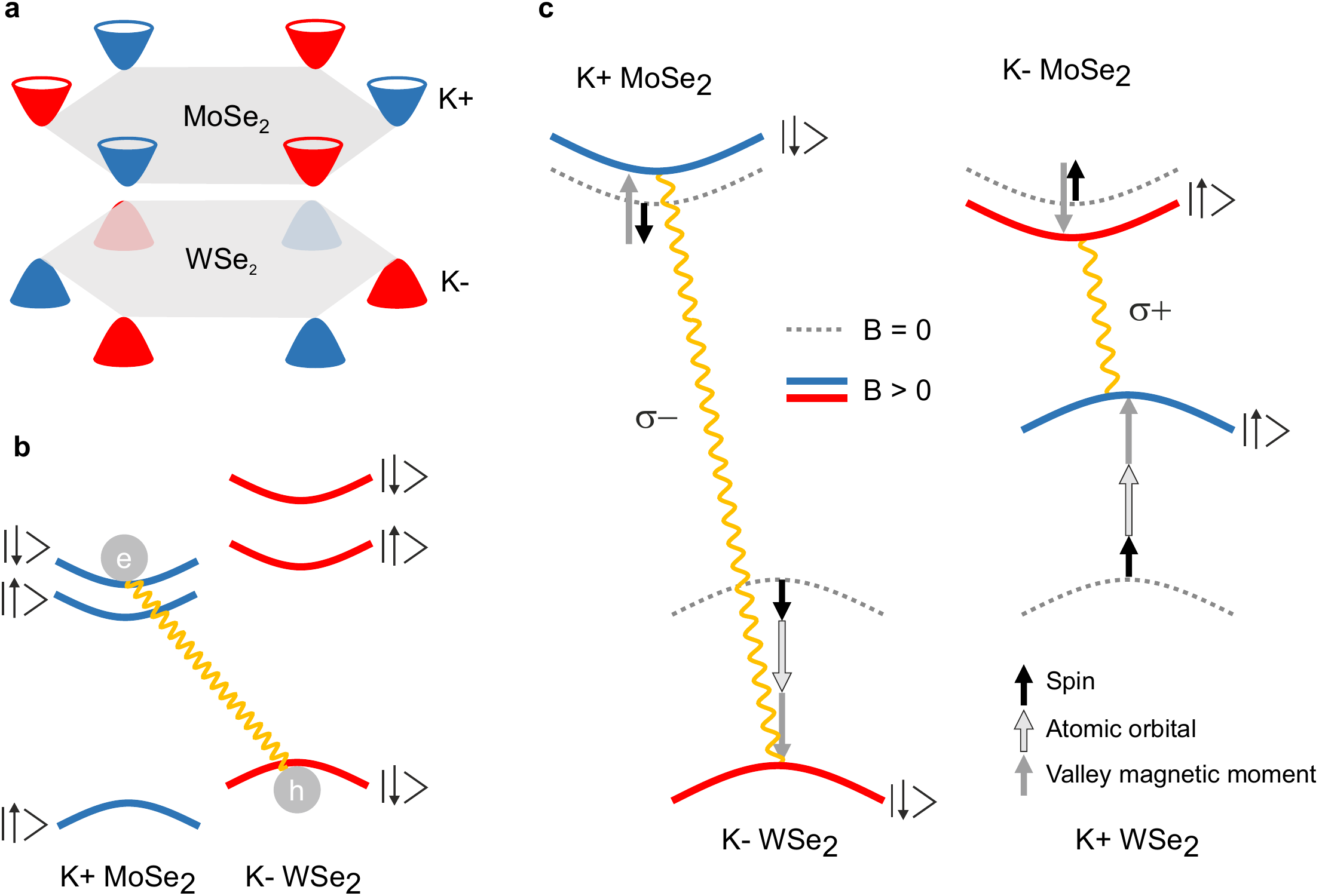}
	\caption{\textbf{Giant Valley Zeeman splitting in an AB stacking configuration.}
		\textbf{a,} Momentum space arrangement of the relevant band extrema of interlayer transitions in an AB-stacked WSe$_2$/MoSe$_2$ heterostructure. Blue (red) depicts electronic bands from the K+ (K$-$) valleys. \textbf{b,} Type-II band alignment of MoSe$_2$ and WSe$_2$ for the AB stacking configuration, indicating spin-allowed optically bright interlayer transitions. Arrows indicate spin-up and spin-down states.  \textbf{c,} Evolution of the energy levels in \textbf{b,} with positive applied magnetic field. Dashed lines indicate the situation for B\,=\,0. The arrows depict the possible contributions to the Zeeman splitting (Black for spin, black framed for atomic orbital and grey for valley magnetic moment).
	}
	\label{Plot3}
\end{figure}

In an AB-stacked heterostructure, however, we encounter a markedly different situation for optically bright transitions. Figure\,\ref{Plot3}a schematically depicts the configuration of the Brillouin zones for interlayer excitons in an AB-stacked WSe$_2$/MoSe$_2$ heterostructure. Here, the optical transitions take place between the K$-$ valley of WSe$_2$ and the K+ valley of MoSe$_2$ (and by symmetry, also from K+ in WSe$_2$ to K$-$ in MoSe$_2$). Hence, in contrast to monolayer systems, the optical transitions are not valley-conserving for AB-stacked heterobilayers. This is further illustrated in Fig.\,\ref{Plot3}b which shows one of the two interlayer transitions.
After the optical excitation and following fast charge transfer, electrons in the upper conduction band of MoSe$_2$ reside in the K$+$ valley whereas the holes in WSe$_2$ are located in the K$-$ valley (here, we use the convention where the vacant electron states and the corresponding holes in the valence band are defined to have the same momenta). This configuration, in analogy to tungsten-based monolayer TMDCs, leads to the spin-allowed and optically bright transitions involving the upper conduction band of MoSe$_2$ while a transition from the lower conduction band is not spin-conserving and thus optically dark. The anomalous situation in momentum space directly impacts the valley splitting of the interlayer exciton in an external magnetic field, which is schematically depicted in Fig.\,\ref{Plot3}c. In analogy to the monolayer system, the contribution to the Zeeman effect from spin cancels out and the magnetic moment from the atomic orbitals $\mu_l$ in the valence bands results in an expected energetic splitting of $-4\mu_B B$.
The contributions from the valley magnetic moment $\mu_k$, however, now have the opposite sign for the conduction and valence bands and thus add up instead of cancelling out as it is the case for individual monolayers. This signifies the main magnetic property of the interlayer transitions in AB-stacked heterostructures with the energetic shifts of the valley magnetic moments evolving anti-parallel in the conduction and valence bands in a magnetic field.
For the $\sigma -$ transition in Fig.\,\ref{Plot3}c this leads to a drastic increase of the transition energy when a magnetic field is applied. For the same reasons, a strong decrease occurs for the  $\sigma + $ transition, since the two valley configurations are linked by the time reversal symmetry.
The total Zeeman splitting of the interlayer transition $\Delta E^{IEX}$ induced by the magnetic field then amounts to $\Delta E^{IEX}=E^{\sigma +}-E^{\sigma -}=-(4+2(m_0/m_e+m_0/m_h)) \mu_B B$.
Using recently calculated values \cite{Kormanyos2015a} for the effective masses of electrons ($m_e=0.57\,m_0$) in the upper conduction band of MoSe$_2$ and holes ($m_h=0.36\,m_0$) in the valence band of WSe$_2$ we obtain a $g$ factor of $-13.1$ for the interlayer transition, in close qualitative agreement with the experimentally determined value of $-15.1\pm0.1$.
\newpage
The presence of large band offsets in TMDC heterostructures resulting in charge transfer across the interface combined with the angle alignment of the nonequivalent valleys thus yields interlayer transitions with an unusually large total $g$ factor. Even higher $g$ factors can be expected when suitable van der Waals materials with lower effective masses are combined. Thus, even for non-selective injection with respect to spin-valley degrees of freedom, strong spin-valley polarization in an applied magnetic field is expected already at thermal equilibrium steady-state conditions, as demonstrated in our experiment. This finding is of particular importance when above bandgap optical excitation or electrical injection are considered, both being most common scenarios in devices. Overall, the experimentally demonstrated field-induced spin-valley polarization in artificial van der Waals heterostructures under magnetic fields illustrates the potential for highly effective manipulation of long-lived interlayer excitons and highlights both the intriguing aspects of novel pre-designed atomically-thin systems and their promise for future valleytronic applications.

\clearpage

\newpage
\section*{Methods}
\textbf{Sample fabrication}
\newline
The WSe$_2$/MoSe$_2$ heterostructure was fabricated by means of an all-dry transfer procedure \cite{Castellanos-Gomez2014a} on a Si/SiO$_2$ substrate. The constituent monolayer crystals (HQGraphene) were obtained by mechanical exfoliation. After the transfer process, the sample was annealed for 5 hours at 150$\,^{\circ}$C in high vacuum.

\textbf{Second-harmonic generation spectroscopy}
\newline
SHG measurements were carried out at room temperature with a Ti:sapphire laser (pulse length 100\,fs, central wavelength 810\,nm) focused on the sample via a 40x microscope objective. The signal was coupled into a grating spectrometer and detected with a CCD camera. For polarization-dependent measurements, the laser light was linearly polarized and the reflected light was analyzed by the same polarizer, thereby selecting the parallel signal component of the SHG. The sample was rotated by a mechanical stage in order to obtain angle resolution.
For mapping of the total SHG intensity, the sample was excited using circularly polarized light without any polarization analysis in the detection. The sample was scanned under the microscope using a motorized x-y stage and the total SHG intensity was recorded for each sample position.

 \textbf{Magneto-PL spectroscopy}
\newline
The sample was placed on a x-y-z piezoelectric stage and cooled down to 4.2\,K in a cryostat filled with liquid helium. Magnetic fields up to 30\,T were applied by means of a resistive magnet in Faraday configuration. For static PL measurements, laser light at an energy of 1.94\,eV was focused onto the sample with a microscope objective resulting in a spot size of $\sim$4\,$\mu$m. The polarization of the PL was analyzed with a quarter-wave plate and a linear polarizer.  Using a nonpolarizing beam splitter, the backscattered PL was guided to the spectrometer and detected with a liquid-nitrogen-cooled CCD. Time-resolved PL measurements were carried out with a pulsed diode laser (laser energy 1.80\,eV, repitition rate 2.5\,Mhz) which was synchronized to an avalanche photodiode. The PL from the interlayer exciton was spectrally selected with a longpass filter.

\section*{Acknowledgements}
 Financial support by the DFG via GRK 1570, KO3612/1-1, SFB 689 and CH 1672/1-1 and support of HFML-RU/FOM, member of the European Magnetic Field Laboratory (EMFL) is gratefully acknowledged.

\section*{Competing financial interests}
The authors declare no competing financial interests.
\end{document}